 \documentclass[final,3p,times,twocolumn]{elsarticle}

\usepackage{amssymb}
\usepackage{xcolor}
\usepackage{soul}
\usepackage{amsmath}
\usepackage{svg}

\usepackage{hyperref}
\usepackage{graphicx}
\usepackage{mathtools}
\usepackage{siunitx}
\usepackage{subfig}
\usepackage{amssymb}

\journal{Powder Technology}

\begin{document}

\begin{frontmatter}



\title{Shear profile in a dense packing of large grains}


\author{Alessio Quaresima$^a$}
\affiliation{organization={Neurobiology of Language Department, Max Planck Institute for Psycholinguistics}, 
            addressline={Wundtlaan 1}, 
            city={Nijmegen},
            postcode={6525 XD}, 
            country={The Netherlands}}
\author{Andrea Plati$^{b,c,d}$}
\affiliation{organization={Université Paris-Saclay, CNRS, Laboratoire de Physique des Solides},
            addressline={1 rue Nicolas Appert}, 
            city={Orsay},
            postcode={91405}, 
            country={Italy}}
\affiliation{organization={Institute for Complex Systems - CNR},
            addressline={P.le Aldo Moro 2}, 
            city={Rome},
            postcode={00185}, 
            country={Italy}} 
\affiliation{organization={Department of Physics, University of Rome Sapienza},
            addressline={P.le Aldo Moro 2}, 
            city={Rome},
            postcode={00185}, 
            country={Italy}}
           
\author{Andrea Gnoli$^{c,d}$}          

\author{Alberto Petri$^{c,d,e}$}
\affiliation{organization={Enrico Fermi Research Center (CREF)},
addressline={via Panisperna 89}, 
            city={Rome},
            postcode={00184}, 
            country={Italy}} 

\begin{abstract}
We investigate the shear of a dense  bed of refracting supermillimetric grains confined within a transparent horizontal annular cell with a rotating top.  The  local time correlation functions of interferometric images allow to characterize the shear profile close to the wall, with a spatial resolution well below the grain diameter. For increasing shear,  we observe  a transition in the system response and the manifestations of anisotropies in the force chains. The employed technique is of easy implementation and is especially suitable for the study of stationary processes.
\end{abstract}

%Graphical abstract
%\includegraphics{GraphAbsCiamb.svg}
%includepackage{svg}


\begin{keyword}
\textit{Sheared granular bed \sep Light scattering \sep Shear band \sep Interferometry}




\end{keyword}

\end{frontmatter}


\section{Introduction}

The response of a granular medium to shear stress is an important and not completely known property \cite{Pouliquen2008}, with many practical consequences in natural phenomena and industry. Different behaviors can be observed, that depend on many variables \cite{GDR2003}. Besides the shear rate itself, these variables  include  grain geometry and tribological properties,  preparation,  volume fraction,   boundary conditions (such as fixed volume or fixed normal stress, etc.)  and possibly others, through dependence that are not always well-understood. Not only the strain rate profile can change, but  highly nonlinear phenomena with consequent instabilities can take place in different circumstances,  like the formation of shear sub-bands \cite{Shukla2019}, stick-slip \cite{Dalton2005}, etc.   Accessing the material bulk in 3D to measure the strain field is not impossible, but certainly hard and resources demanding  (see e.g. \cite{Yang2002,Amon2017a,Fabich2018} end refs. therein). It can therefore be  useful to have at least a proxy \cite{Bocquet2001} 
of how the granular medium deforms and flows under shear in the different regions observing what happens at the medium boundaries.

A tool that has proven particularly fruitful for the study of granular dynamics is the scattering of light. Depending on the system and the phenomenon at hand, different techniques have been developed.
One widely employed is Diffusing Wave Spectroscopy (DWS) \cite{Weitz1993}. It  requires  a transparent medium, in which  coherent
 light can undergo multiple scattering  before  reaching the detector, giving rise to a diffusive process generating a grainy pattern of speckles.
 DWS can be successfully employed  to study the dynamics of both submicrometer particles \cite{Weitz1993} and of larger, submillimeter, particles \cite{Amon2017}, yielding information on the  motion of grains within a certain depth 
from the system boundaries.  In  larger particles, the diffusive process is generated by multiple reflections and refraction at the grain interfaces.

In the case of supermillimetric  particles, absorption prevents light from being scattered a large number of times and from entering the diffusive regime. In addition,  the grain size can be comparable with  the  coherence length of the laser beam. Thus, different methods have been adopted that allow  
the study of grains adjacent to a transparent wall 
or to the system surface. Among the most used  are Particle Tracking Velocimetry (PTV) and Particle Imaging Velocimetry
(PIV) \cite{Hagemeier2015}. PTV consists of directly tracking  the particle trajectories  of the grains. This technique can in principle attain high precision, however, it requires that grains remain always visible, which  may be not verified in 3D systems, where they can migrate from the boundary to the bulk and vice versa, and hence PTV is generally not suitable for long-time measurements.
PIV is based on the comparison of highly contrasted images taken at different times. Deformation is evaluated by optimizing the overlap of the images, which can be quite demanding \cite{Sarno2019}. PIV needs very bright images, often obtained with the use of very powerful light sources  ($\simeq 0.5$ kW) and a mix of well-contrasting beads, e.g. black and white. Moreover, image processing  requires    
elaborated software, being performed usually  in the Fourier space. At the same time, for inhomogeneous systems, images must be split into several windows to get the desired spatial resolution, with drawbacks on the Fourier Transform. Another technique widely used is the Digital Image Correlation (DIC) \cite{Lei2018}
 which can give detailed results on the strain field after preparing the medium to yield highly contrasted images, as in PTV.    

 In this  paper, we propose an experimental procedure, to the best of our knowledge original, to investigate deformations and flows at the boundary of a dense assembly of refracting grains.
  The investigated system consists of glass beads  contained in a horizontal transparent channel and sheared at the top. 
 The procedure  exploits the interference pattern generated  by reflection and refraction at the air-grain interface and possibly by scattering from rough grains 
 \cite{Dossow2021}, and is sensible to the motion of particles closer to the container wall.  
 Information on the vertical profile of the shear band is obtained from the time correlation of the light intensity.  
 The advantage of the technique lays in its simple implementation and in its ability  in resolving the dynamics well below the grain size, yielding spatially detailed information.

The aim of this work is two-fold. On the one hand to show the effectiveness of the proposed technique. On the other hand to present the main, specific, features characterizing the shear band in the investigated system. 
In Sec.~2, we describe the experimental setup and the data processing.  The main results are reported in Sec.~3 while Sec.~4 contains a discussion and a summary of the main findings.  Several technical and methodological details can be found in the Supplementary Material (SM), together with the URL  where an illustrative movie is stored, and  a Julia  code  expressly developed to process the images is made available.

\section{Material and methods}

%\subsection{Metapixel}

%The simultaneous acquisition of intensity values at many  different points allows to compute averages \cite{Viasnoff2002}
%This technique can be easily performed using of  arrays of synchronized detectors, like active pixel sensors. Moreover, performing ensemble averages, it can also be usefully employed  for the study of slow dynamics and non-stationary processes.  

\subsection{Experimental apparatus}

The granular system under investigation  
%(Fig. \ref{fig:setup}) 
is similar to the one  described in~\cite{Baldassarri2019} (see SM for details).  
\begin{figure}
\centering
\includegraphics[width=0.995\columnwidth]{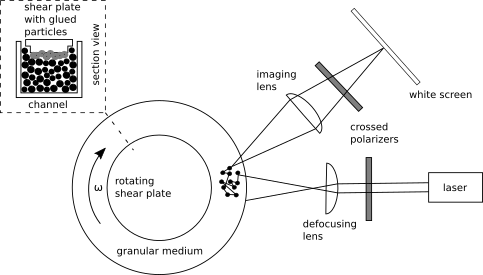}
\caption{ %Top: draft-figures-1
Schematic sketch of the  experiment.}
\label{fig:setup}
\end{figure}
Grains are confined in a transparent horizontal annular channel with inner and outer radii $r=12.5$ cm and $R=19.2$ cm, respectively. To prevent crystallization, it is composed of a $50\%-50\%$ mixture of hard glass beads, 1.5 and 2.0 mm in diameter. An aluminum plate lays on the top, in direct contact with the granular material, and is connected to a motor so that it can shear the system at constant angular speed $\omega$.
A layer of grains is glued to its lower side in order to increase drag. However, it does not extend to the whole plate surface, but excludes an annulus close to the border, to prevent free grains from getting stuck between glued grains and container walls (see the section view in Fig.
\ref{fig:setup}). It will be seen that this mechanical feature is revealed by optical measurements.
As sketched in Fig. \ref{fig:setup},  interferometry is performed in backscattering through a single mode Nd:Yag laser yielding  elliptically polarized light  at a wavelength of 532 nm,  with a maximum nominal power  of 500 mW.
The horizontal laser beam is filtered by a polarizer and is expanded  by a diverging lens. 
The back-scattered light is then focused towards a perpendicular white screen by another lens and eventually passes through a crossed polarizer, to minimize the amount of light directly reflected by the surfaces of the container wall.
The resulting interference image is framed by a digital camera equipped with a CMOS array, at a  rate of 15 frames per second. 
The size of the probed region is limited by at least three factors: \textit{a)} the laser power,  which restricts the scattering area and the thickness of the granular medium involved. With the employed equipment, a sufficient intensity contrast is still obtained with a spot of about 20 mm in linear size, i.e.  10-13 bead diameters, independently of the chosen optical magnification for the image; 
 \textit{b)} the coherence length of the beam, which determines the optical depth of the region where interference can be generated; \textit{c)} the curvature of the container, which gives rise to optical aberration and consequently requires a sufficient narrow  imaged area. In order to get a good focusing, the image has been cropped 
 horizontally. 

\subsection{Data processing and analysis: coarse-graining and noise reduction}

\begin{figure}
\centering
\includegraphics[width=0.99\columnwidth]{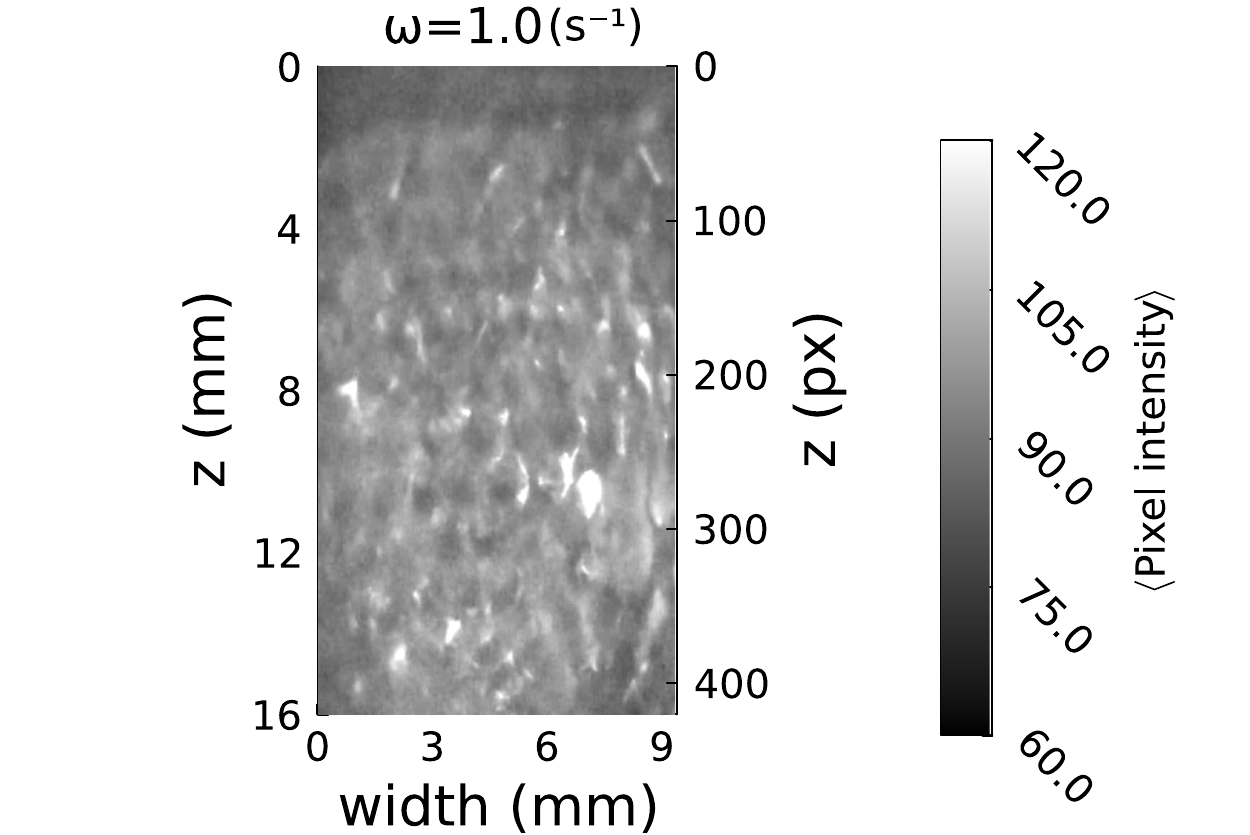}	
\caption{An interference image showing the spatial reference system of our setup. The position of the rotating plate coincides with $z=0$.}
\label{fig:interf}
\end{figure}

The size of the interference pattern depends on the width of the laser spot and on the optical magnification  $M$ of  the scattered image, that can be chosen opportunely.   An example of the emerging pattern is shown in Fig.~\ref{fig:interf}.  The image is made of 240 (horizontal) x 420 (vertical) pixels whose intensities have been normalized as explained in the SM. One pixel corresponds to a region of linear size 0.04 mm on the container wall. The surface of the material in contact with the rotating plate coincides with the top of the image ($z=0$). 
To detect how  the granular system deforms under different shear rates at different depths, video records of the interference pattern have been shot at a rate of 15 frames per second for  about 11 minutes  (see SM for more details).  So, each video supplies about 10$^5$ time series of 9.9k points each. 
\begin{table}[h]
	\centering
		\begin{tabular}{|l|c|c|c|c|c|c|c|c|}
			 \hline
			\label{speeds}
			
$\omega$  (s$^{-1}$)      & 0.05 & 0.07  &  0.1  &  0.5  & 0.7   & 1.0   & 1.2   & 1.5   \\
			\hline
$v$ (mm   s$^{-1}$)     		  & 9.6   & 13.4  & 19.2  &  96  & 134  &  192 &  230 & 288   \\
		\hline
		\end{tabular}
		\caption{Values of plate speed considered in the experiments in radiant per  second. The lower line reports the plate speed  in mm s$^{-1}$ 
       at the outer wall of the container.}
\end{table}
 However, the signal 
 obtained from single pixels are generally highly noisy. For this reason, it is often convenient to perform  an average of the intensity over larger regions, covering $S^2$ pixels, also called \textit{metapixels} \cite{Viasnoff2002}, chosen on the base of the experimental conditions. In the present case 
 the size $S$  of the metapixel  has been chosen through a procedure (detailed in the SM) aimed at getting a trade-off between spatial resolution and intrinsic noise, yielding  $S=10$ pixels,   corresponding  to  $\approx 1/4 \div 1/2$  bead diameters. Thus, the final interference images are composed of 24 (horizontal) x 42 (vertical) metapixels. Each metapixel is identified by an integer horizontal coordinate $x \in [1,24]$, from left to right, and a vertical coordinate  $z \in [1,42]$, from top to bottom.
 
For every value of the plate speed reported in Tab.~1 the  instantaneous intensity on each metapixel, $I(x,z,t)$  has been obtained averaging the intensities on the corresponding metapixels; 
Then, the connected autocorrelation functions have been computed on each metapixel as a function of position and time; in the following equation, the variables \textit{x,z} are omitted for clarity, i.e.,  $I(x,z,t) = I(t)$.
\begin{eqnarray}
&g_\omega(x,z,t)=\cfrac{\langle I(t_0) I(t_0+t) \rangle_{t_0} - \langle I(t_0)\rangle^{2}_{t_0}}
{\langle I^2(t_0)\rangle_{t_0} - \langle I(t_0) \rangle^{2}_{t_0}}&
\label{corr}
\end{eqnarray}
where, as indicated by the subscript, averages are made over different initial times $t_0$. Their  values at discrete times   $t = n~\cdot~\Delta t$ with $\Delta t =1/15$ $s$,  have been obtained through the corresponding empirical estimates (see SM).  
A purposely devised code with a parallelized structure has been developed to process the data\footnote{The code is made available for public use, see on-line SM}.
\begin{figure}
	\centering
	\includegraphics[width=0.99\columnwidth]{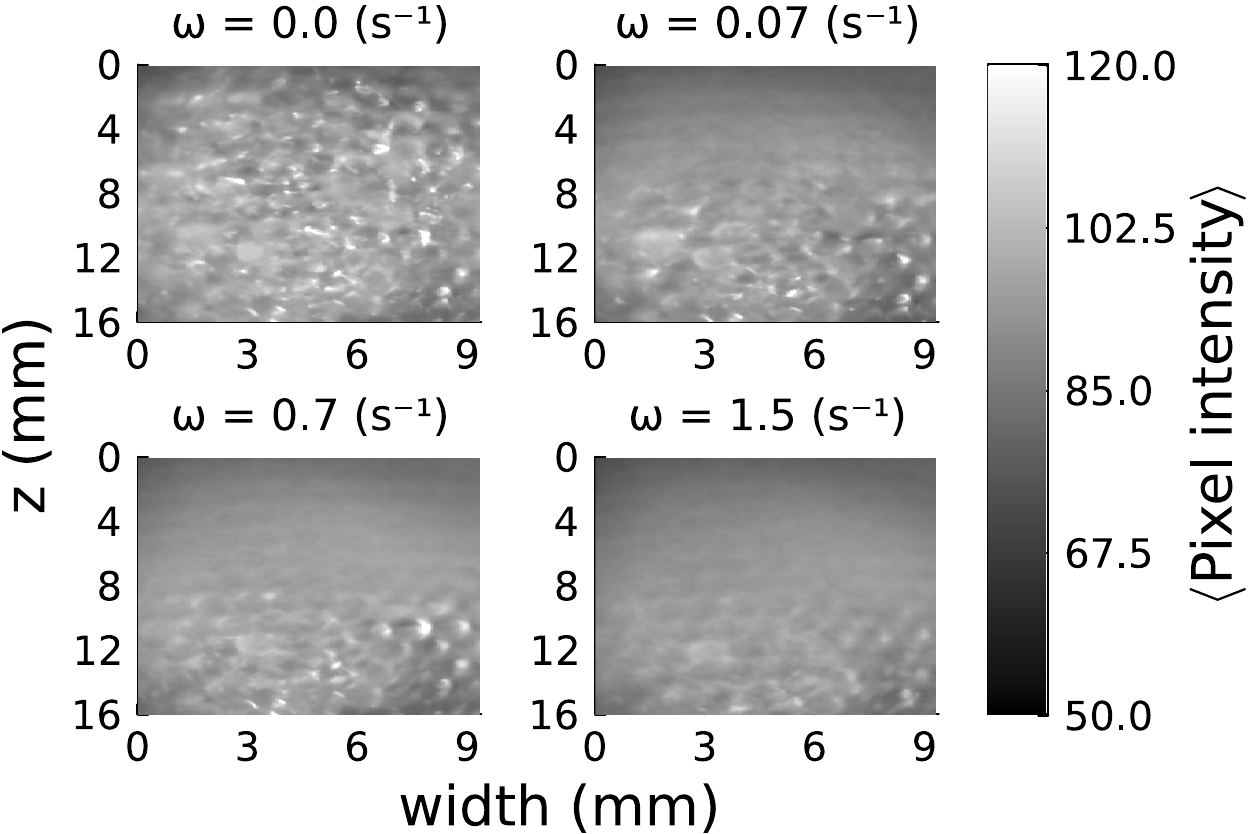}	
	\caption{Intensity matrices averaged over time for the entire duration of the experiments. Images become more blurred for increasing plate  speed (from top left to bottom right). }
		\label{fig:compare}
\end{figure}

\section{Results}

\subsection{Phenomenology}
 
Before discussing autocorrelation functions, we first qualitatively characterize the phenomenology of the system by looking at the time-averaged intensity matrix. 
We recall that different mechanisms may contribute to the time evolution of scattered light intensity. On the one hand, grains slowly diffuse, changing their relative position and thence the interference pattern generated by refractions and reflections at the glass-air interfaces.  On the other hand, each single grain moves as a rigid body and can originate changes in the pattern due to the light scattering occurring at its rough surface   \cite{Dossow2021}. 

Figure \ref{fig:compare} shows the time average of the intensity matrix for some different velocities of the driving  plate. Such an average has been performed over the entire experiment duration producing the same effect of a high exposure time in a photo. Indeed, the image appears blurred in the regions where intensity changes in time \footnote{This effect is exploited in the so-called Speckle Visibility Spectroscopy
or Single Exposure Speckle Photography}. 
In the uppermost part  (closer to the rotating plate) the image is blurred rather homogeneously, while in the bottom area, an intensity pattern can  be neatly seen as for the system at rest.  Homogeneously blurred regions reflect a fast and disordered motion of the particles while spotted areas contain grains that are almost still during the experiment.   As expected, the region in which flowing motion is visible increases in size for increasing shear velocity,  extending deeper.  This allows to discriminate among  different behaviours 
of the interference figure at different depths and to identify a  shear band. 
An illustrative movie showing intensity over the entire time duration of the experiment is linked to the on-line SM.

 \begin{figure}
	\centering
		\includegraphics[width=0.99\columnwidth]{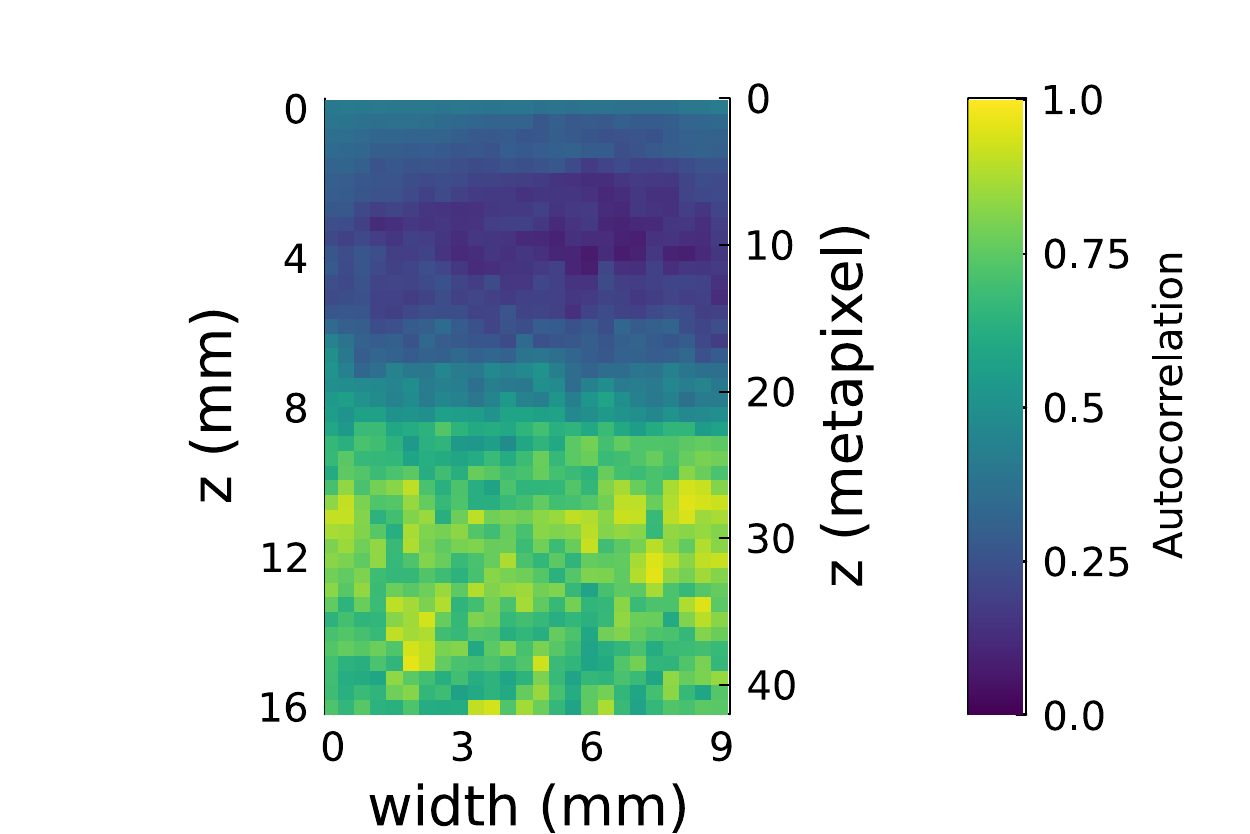}
\caption{Autocorrelation value after 15 frames (1 second) for each metapixel, for plate speed $\omega=1$ $s^{-1}$.}
		\label{fig:allmetapix}
\end{figure}

\subsection{Shear band characterization}

\subsubsection{Autocorrelation decay}

The above considerations can be made more quantitative by analyzing the time behavior of the intensity autocorrelation Eq. (\ref{corr})  for different depth $z$ and plate speed $\omega$. 
Figure \ref{fig:allmetapix} shows an example ($\omega=1$ $s^{-1}$) of how fast intensity decorrelates at each meta-pixel. The color map encodes the autocorrelation  values $g_{\omega}(x,z,T)$ after $T = 1$ second (15 frames). %that is $g_{\omega}(x,z,T)$$. 
The map shows that the metapixels decorrelation speed depends on the depth but not on the horizontal position, as expected from  the symmetry of the system (see also the illustrative movie linked to the SM).

\begin{figure}
	\centering
	\includegraphics[width=0.99\columnwidth]{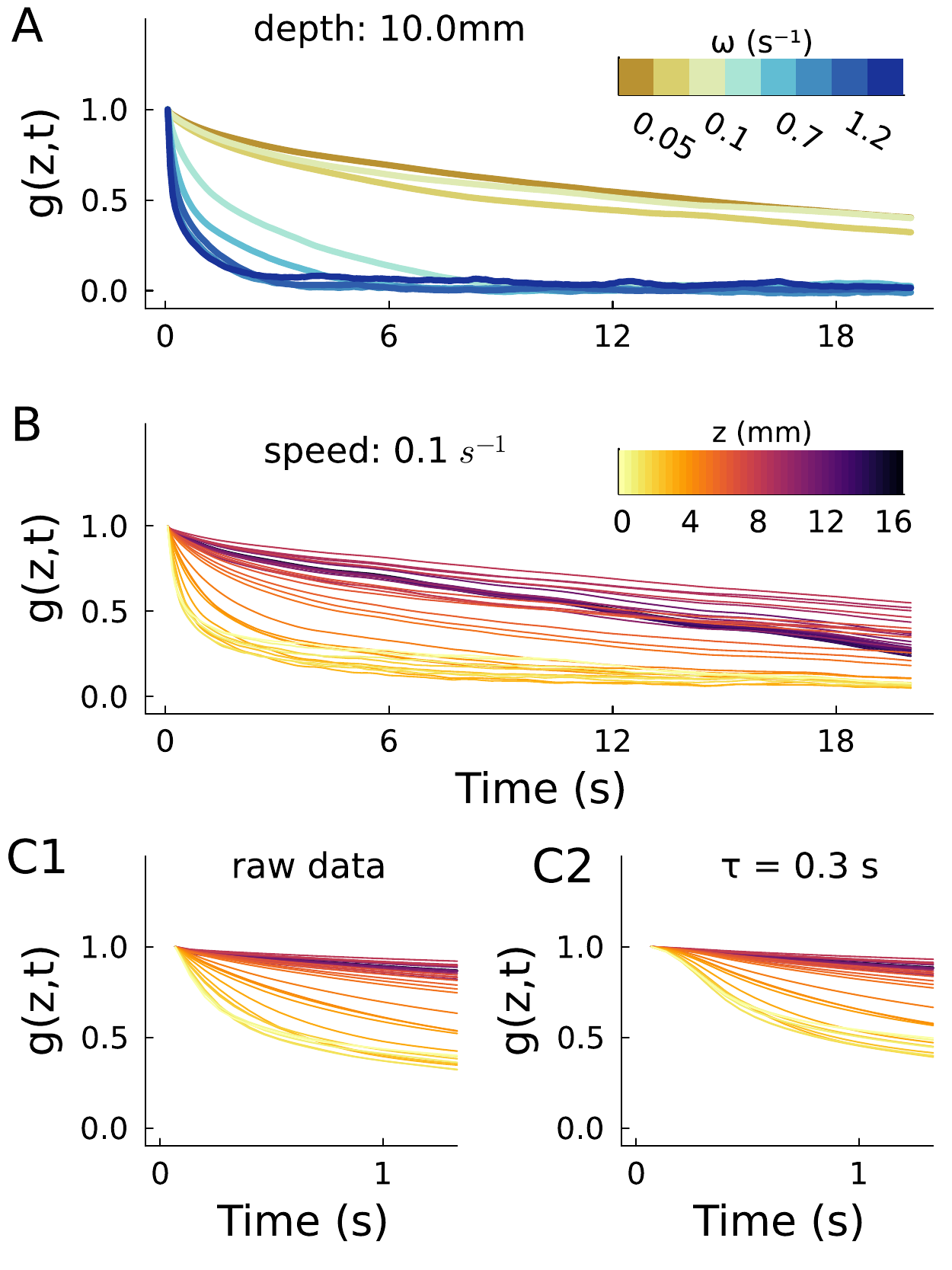}
	\caption{A) Autocorrelation function for different driving velocities $\omega$ at a depth $z=10$ mm. Curves are shown without statistical errors for clarity. B) Autocorrelation functions at different depths $z$ and fixed $\omega=0.1$ s$^{-1}$. C): Comparison between autocorrelations of raw intensities (C1) and the ones obtained from the same signals after a running average over 5 frames $\simeq$ 0.3 seconds (C2).} \label{fig:gdiffv}
\end{figure}

In the following analysis, being the system rotationally invariant, we have considered the curves $g_{\omega}(z,t)$
obtained by averaging $g_{\omega}(x,z,t)$ over $x$. 
Examples of the curves obtained in this way for different speeds are shown in Fig.~\ref{fig:gdiffv}A
  at a fixed depth $z=10$ mm. As expected, the decay is faster at larger speeds.
 In a similar fashion, the autocorrelation at different depths slows down for increasing distance from the plate. Examples are shown in 
 Fig.~\ref{fig:gdiffv}B, 
 for $\omega=  0.1 $ s$^{-1}$. 

\subsubsection{Shear profile: Best fit estimates}

\begin{figure}
	\centering
  		\includegraphics[width=0.99\columnwidth]{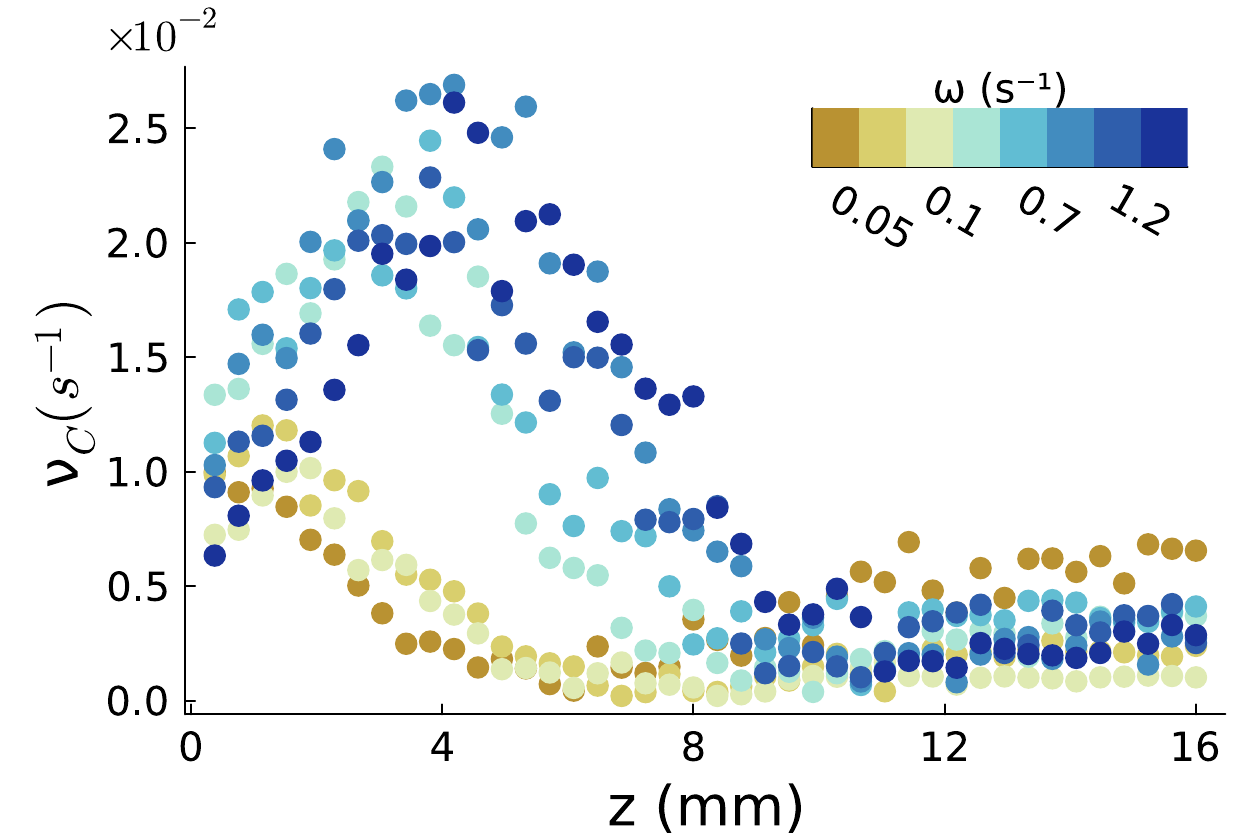}
\caption{Characteristic inverse times $\nu_C=\tau_C^{-1}$, representative of the granular strain rate from Gaussian fits of smoothed autocorrelations at short times for the different $\omega$ after averaging on the horizontal coordinate.}
		\label{fig:gausspar}
\end{figure}

In order to characterize the shear profile we performed a more detailed analysis of the autocorrelation decay. To this aim,  we estimated a characteristic decay time $\tau_C(z)$ by fitting the initial part of $g_\omega(z,t)$. Considering that the decorrelation of the interference pattern is caused by grains motion, we can use $\tau_C(z)$  as a proxy of the typical time needed for the system to perform a local spatial rearrangement at depth $z$. Then, the shear profile can be defined through the decorrelation rate $\nu_C(z)=\tau^{-1}_C(z)$.     
Since the finite acquisition rate (i.e.~15 fps) can cause abrupt decorrelations at short times, the fitting procedure has been performed on autocorrelation functions of smoothed intensity signals. Such smoothing has been done with a running average over a time window $\tau$.  Examples of autocorrelation functions  obtained from smoothed signals are reported in the bottom panel of figure \ref{fig:gdiffv}C. Their decay at short times can be  fitted by  a Gaussian function $ \propto \exp(-(t/\tau_G(z))^2)$. However, the fitted time $\tau_G(z)$ is not equivalent to the physical decorrelation time $\tau_C (z)$ since the short time behavior of the autocorrelation functions is affected by the smoothing procedure (which in turn introduces an additional timescale $\tau$). The analytical expression of the autocorrelation function obtained from a smoothed signal reveals that the physical decay time depends on $\tau_G(z)$ and $\tau$ through the relation $\tau_C(z)=2\tau^2_G(z)/\tau$ (details of the calculations are reported in the SM). Then, fitting $\tau_G(z)$ and knowing $\tau$ allows to estimate $\tau_C(z)$. 
The shear profiles $\nu_C(z)=\tau_C^{-1}(z)$ obtained from autocorrelation functions of signals smoothed over $\tau=0.33$ s (5 frames) are reported in Fig. \ref{fig:gausspar}. As expected, $\nu_C$ at a fixed depth displays larger values at higher shear rates. 

A first remarkable result is that the technique is able to resolve the dynamics well below the grain size. In fact, while the image of one grain spans 2-4 metapixels (see SM), $\nu_C$ is seen to significantly change on the scale of the single metapixel.
It is also interesting to note that, for any imposed shear velocity $\omega$,  the decorrelation rate $\nu_C$ stays finite at all the observed distances from the shearing plate, being generally larger for larger $\omega$. Therefore, although very slowly, the grains flow at any depth. This makes it difficult to define the shear band in a strict manner, however, it is seen that below the surface $\nu_C$ displays a local maximum, after which it decays until an inflection point. One can  conventionally identify this point with the shear band limit $Z_B$ (see SM). 
\begin{figure}
	\centering
	\includegraphics[width=0.99\columnwidth]{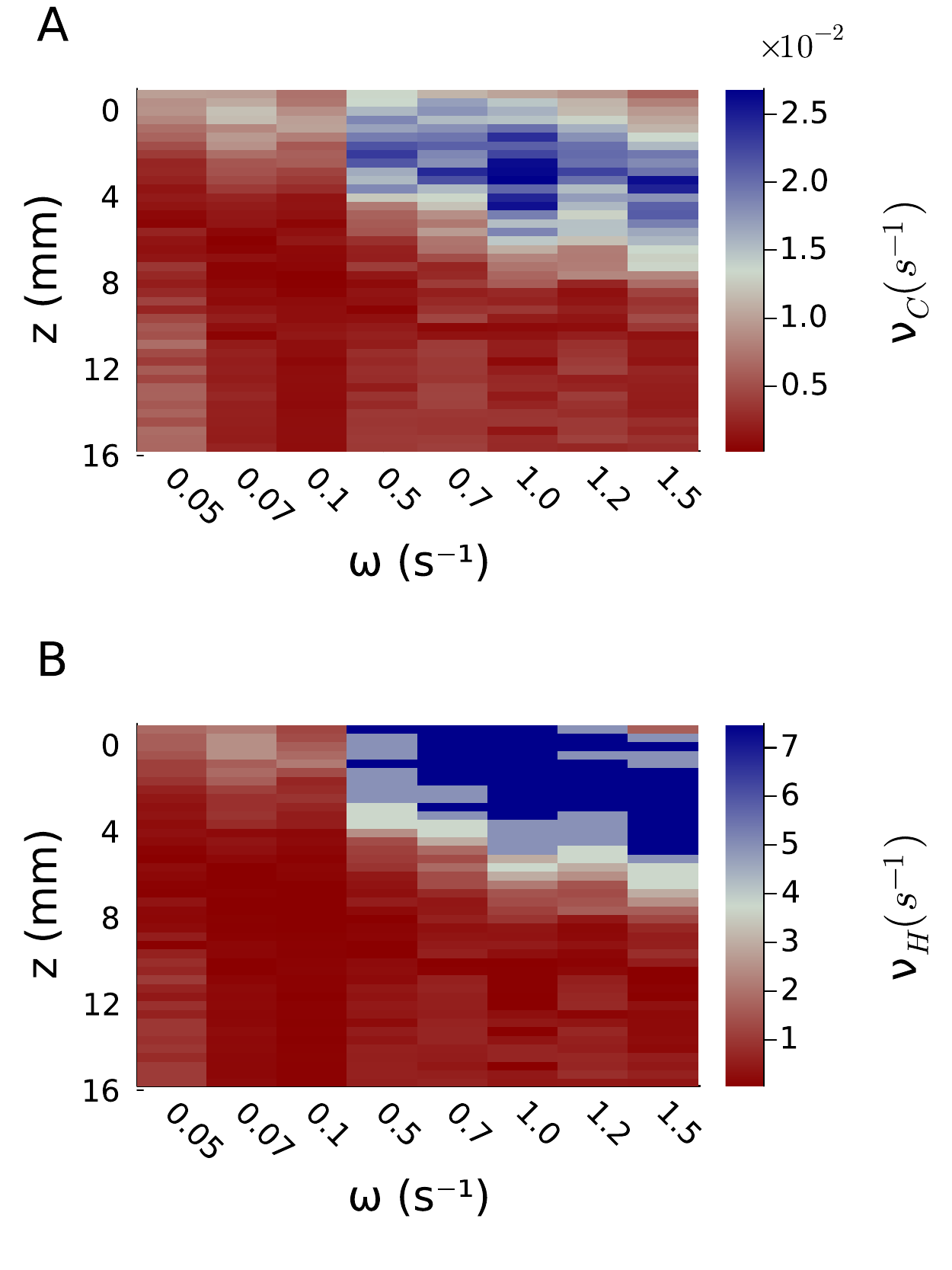}
	\caption{A)  Decorrelation rates from Gaussian fits $\nu_C$ at different depths and plate speeds. B) Half-height decorrelation rates $\nu_H$ at different depths and plate speeds.}
	\label{fig:SBcolor}
\end{figure}
Another feature displayed by the different curves in Fig.~\ref{fig:gausspar}
 is that $\nu_C$ does not attain
its maximum $\nu_C^{max}$  just below the plate but at a certain depth $Z_{max}$.   This  can be understood considering that, as described in \S2.1, the lower side of the plate,  in contact with the grains, is smoother close to the container walls, and therefore its dragging is less effective at the boundary  than in the bulk.  The appearance of a maximum shear at a certain distance from the plate is consistent with the transmission of the shear stress  through a network of force chains which  results to follow  an average inclined path from the central region of the plate to the wall.
At the same time it may seem weird that, increasing the plate velocity, the region of maximum shear rate deepens, as seen in Fig.~\ref{fig:gausspar}. 
It is known \cite{Behringer2019,Meng2021} that
systems develop anisotropy in the contact  force  network along the direction of the shear so that stress propagates at an average angle with respect to the vertical that increases as a function of the shear rate. Likely,  anisotropy affects  force propagation in all directions and the observed increase of $Z_{max}$ with  $\omega$ suggests that force chains in the  direction transversal to the shear undergo an opposite fate, decreasing their angle with respect to the vertical for increasing shear rate.
We have not been able to find related results in the literature, so this would be the first evidence of such a mechanism, to be corroborated by additional experiments and simulations.

A pictorial view of the shear band is shown in Fig. \ref{fig:SBcolor}A. It consists of a color map of the same $\nu_C$ reported in Fig. \ref{fig:gausspar} after normalizing all values  to their maximum for each plate velocity. The vertical scale reports the deepness $z$.
It can then be noticed that for each speed there is an upper region where  $\nu_C$ is larger and changes more rapidly. This region extends deeper for increasing plate speed 
evidencing a main shear band.

\begin{figure}
	\centering
	\includegraphics[width=0.99\columnwidth]{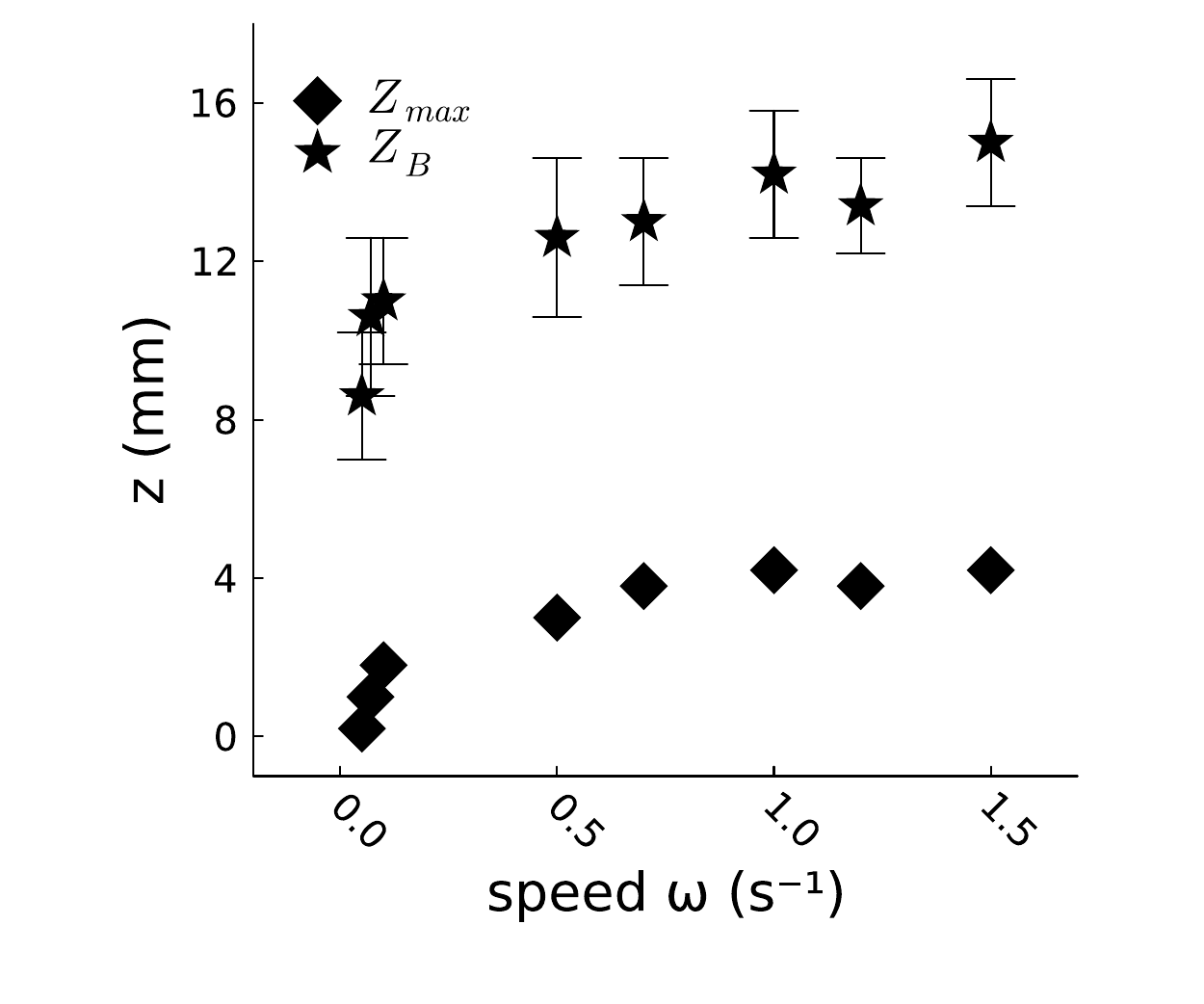}
	\caption{Characterizing quantities for the shear band as functions of $\omega$.
  The automatized procedure to detect such regions is explained in the SM together with the definition of the error bars for $Z_B$.}	\label{fig:gauvaluessinvtimespeak_vs_z}
\end{figure}

 The behavior of the main quantities characterizing the shear band as a function of the plate velocity is  resumed in Fig.~\ref{fig:gauvaluessinvtimespeak_vs_z}. 
 Noticeably,  both quantities display a similar dependence. 
 Two different linear regimes can be envisaged, separated by a crossover velocity $\omega_c \approx 0.1$ s$^{-1}$. In the case of a compliant coupling between the motor and the plate, $\omega_c$ corresponds to a transition from plate  stick-slip  to  continuous sliding \cite{Baldassarri2019}.  This transition does not depend on the stiffness of the coupling \cite{Dalton2005} but is  intrinsic to the medium response to shear in this system. 
Therefore it  gives rise to the appearance of different  regimes  even in the present case of a rigid coupling where the plate cannot perform stick-slip.

%Synthetic
\subsubsection{Shear profile: Fast characterization}
\begin{figure}
	\centering
		\includegraphics[width=0.999\columnwidth]{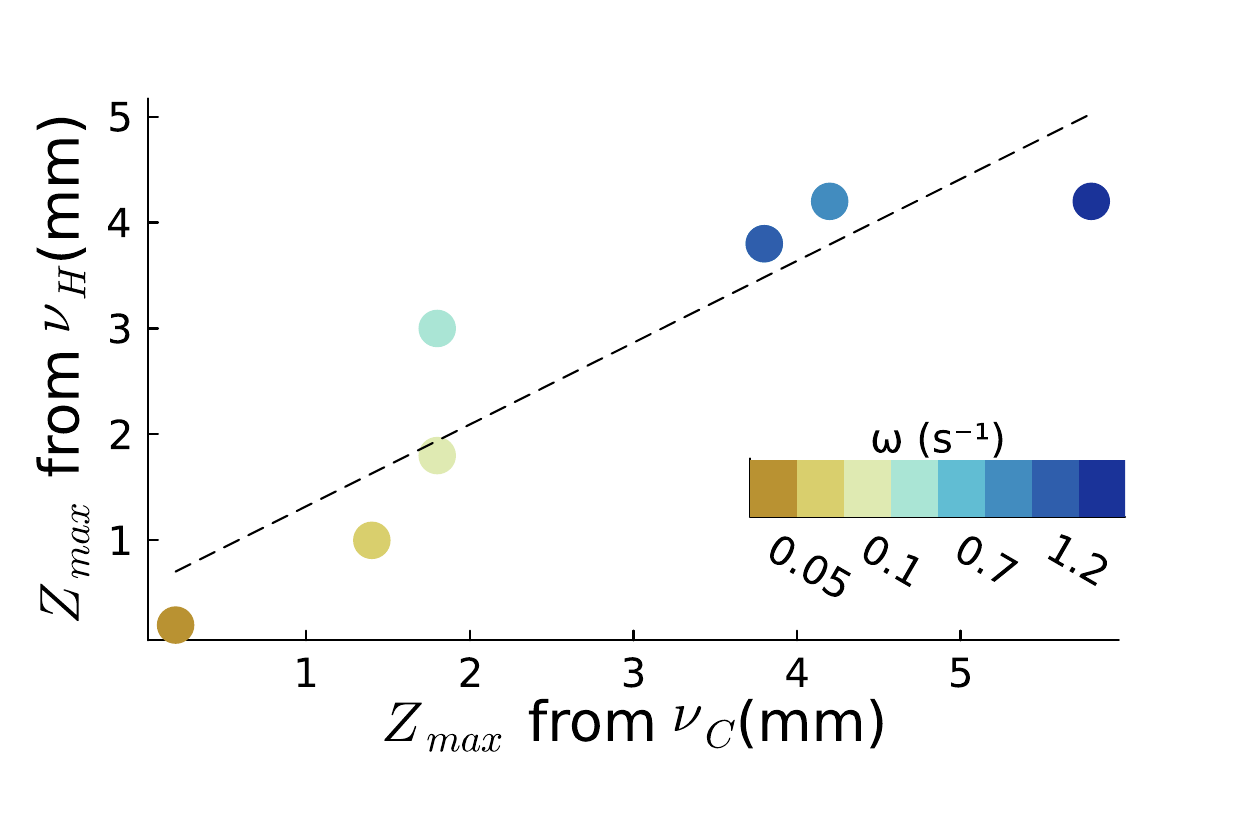}
	\caption{Correlation between the height of the characteristic maxima of decorrelation rates $\nu_H$  and $\nu_C$. }
	\label{fig:cpmtau}
\end{figure}

For many practical applications, it can be  desirable to have a fast way for characterizing the autocorrelation decays, possibly avoiding performing many best fits to all the curves. To this aim, we have evaluated the half-height decorrelation rate $\nu_H=\tau_H^{-1}$, where $\tau_H$ is the time at which the value of  $g_{\omega}(z,t)$  is halved (we point out that $\nu_C$ is obtained from autocorrelations of smoothed signals while $\nu_H$ is not). As a first qualitative comparison with the best fits, we report in Fig. \ref{fig:SBcolor}B the shear-band colormap obtained with $\nu_H$, which exhibits a phenomenology in good agreement with Fig. \ref{fig:SBcolor}A. Indeed, for each omega, $\nu_H$ is able to distinguish between the slow-flowing region at large depth and the fast-flowing one at lower $z$. However, within this latter region, $\nu_H$ reveals to be less accurate than $\nu_C$ because of the finite acquisition rate (see also Fig. S4 in the SM).
 A more quantitative test can be done by comparing the depth of the maximum decorrelation rate obtained from $\nu_H$ with the one obtained from the best-fit parameter $\nu_C$ for each value of $\omega$. Figure \ref{fig:cpmtau} reports  a regression  between the two quantities,  showing a good degree of correlation. 
The half-life times  can therefore be employed  to  characterize the shear band  in a fast and simple way.

\section{Conclusions}

We have experimented an interferometry-based imaging technique to investigate the shear  in a dense packing of supermillimetric transparent grains confined in a transparent channel and sheared by a rotating top lid. The technique is of easy implementation and capable of resolving the shear  profile  well below the grain size. Different characterizing parameters can be extracted from the analysis of  local intensity time autocorrelation  functions. Their similar dependence on the plate speed shows that for the investigated system they are equivalent.  The technique has also allowed to detect the existence of two shear regimes, separated by a crossover which corresponds to the transition from granular stick-slip to continuous sliding.  Shifting  of the maximum shear to larger depths for increasing plate speed reveals the development of anisotropies in the force chains.
Our comparative analysis of the parameters  shows that a quick and robust characterization of the shear profile can also be obtained straightly from the half-life time of the correlation functions. The technique can potentially be employed in different situations, especially for the study of  stationary processes, and more internal layers could  be probed increasing the
coherence length and power of the light source.

 \section{Acknowledgments}
The authors acknowledge  R. Cerbino, M. Leonetti  and
 G. C. Ruocco for  useful discussions. 
\\
\\

This research did not receive any specific grant from funding agencies in the public, commercial, or not-for-profit sectors.
 
\bibliographystyle{unsrt}
\bibliography{reference-3}

\newpage

%\documentclass{article}
%\usepackage{hyperref}
%\usepackage{graphicx}
%\usepackage{mathtools}
%\usepackage{xcolor}
%\usepackage{siunitx}
%\usepackage{subfig}
%\usepackage{amssymb}
%\usepackage{soul}
% $\unit{\frames} = \text{frames}$

%\graphicspath{{fig_supp}}
%\title{}
%\author{}
\onecolumn
%\begin{document}
\section*{Supplemental Material}
%\maketitle

\section*{Experimental setup}
\subsection*{Mechanics}
The mechanical experimental setup used for this research (Fig. S\ref{fig:setup}) is  based on a circular horizontal PPMI channel of outer and inner radii $R$
= 19.2 cm and $r$ = 12.5 cm respectively,  and 12 cm height. A bidisperse mixture of \SI{50}{\%} - \SI{50}{\%}
glass beads, of radii $r_1=$ \SI{1.5}{\milli\meter} $\pm$ \SI{10}{\%} and $r_2=$\SI{2}{\milli\meter} $\pm$ \SI{10}{\%}, fills the channel to about \SI{9}{\centi\meter}  in height. 
A horizontal plate fitting the channel can shear the granular medium from the top. It is free to move vertically and can be rotated. A layer of grains is glued to its lower face in order to better
drag the underlying granular medium. It does not extend to the whole plate surface, but excludes an annulus close to the border,  to prevent free grains from getting stuck between glued grains and container walls (Fig. S\ref{fig:setupSI}). The plate has mass $M = $\SI{1200}{\gram} and moment of inertia $I =$\SI{0.026}{\kilo \gram \meter \squared}, implying that in our experiments the medium is  under a nominal pressure $p=Mg/[\pi(R^2-r^2)] \simeq $ \SI{176}{\pascal}. 
The plate can be rotated through a stepping motor, at constant angular speed  $\omega_p$ in the interval of \SIrange{0.05}{1.5}{\radian/\second}.

\subsection*{Optics}
\begin{figure}[h]
	\centering
	\includegraphics[width=0.8\linewidth]{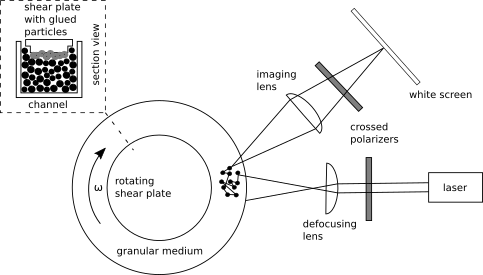}
	\captionsetup{labelformat=empty}
	\caption{Figure S\ref{fig:setupSI}: a) Sketch of the experimental set-up with a  detail of the top plate.} 
	\label{fig:setupSI}
\end{figure}

The optical setup is also sketched in Fig. S\ref{fig:setupSI}. 
A Nd:YAG laser supplies an elliptically polarized light beam at a wavelength of \SI{532}{\nano \meter},  with a maximum nominal power of \SI{500}{\milli\watt}.
The beam is filtered by a polarizing filter and is expanded by a  lens with $f=50 $ mm, hitting the channel wall horizontally.
The back-scattered light is then focused towards a perpendicular white screen by another $f=50$ mm lens. 
To minimize the amount of light reflected by the perspex walls and by the first vertical layer of grains, the beam eventually passes through another polarizer, crossed with the first. 
%the back-scattered light passes through another polarizer, crossed with the first. The beam is eventually focused on a perpendicular white screen by another $f=50$ mm lens. APL

The resulting interference figures have been recorded with a  D90  Nikon camera equipped with a CMOS, of 640$\times$424 pixels in the horizontal and vertical direction respectively, and a dynamics of 8 bits.
A lens with $f = 50$ mm has been employed, with an acquisition rate of \SI{15}{\text{frames}/\second}.  
Due to the channel curvature and the laser spot dimension, the useful area of the interference pattern on the detector had a size of 250$\times$420 pixels. Calibration measures have shown that one pixel corresponds to a region of the size of \SI{0.04}{\milli\meter} on the container wall, so the analyzed region of the granular system has an external surface of \numproduct{10 x 16.8} \unit{\milli \meter \squared}.
The grains diameter is, on average, of about \SI{1.75}{\milli \meter}, corresponding to  \SI{44}{\text{pixels}}.
The main typical linear sizes characterizing our system are resumed in Tab. S1.
\begin{table}[h]
	\begin{centering}
		\begin{tabular}{l|r|l}\label{tabell}
			element         & mm        & pixel       \\
			\hline
			pixel         &   0.04   & 1       \\
			grain         &   1.5 - 2.0   &  38 - 50         \\ 
%   		$l^{*}$           & 5.75     & 150       \\
			interference figure (width $\times$ height)   & \numproduct{10.0 x 16.8}   
			  & \numproduct{250x420}
			  
			 \\
			metapixel & 0.4    & 10\\
%			speckle   &  0.12 $\div$ 0.15 &  3 $\div$ 5
		\end{tabular}
		\captionsetup{labelformat=empty}
		\caption{Table S1: Some sizes characterizing the system}
		\label{tab}
	\end{centering}
\end{table}

\section*{Preprocessing of video}

\begin{figure}
	\centering
%    \subfloat[\centering]{{\includegraphics[width=5.5cm]{SIFig1} }}%
%    \qquad
%    \subfloat[\centering]{{\includegraphics[width=5.5cm]{SIFig2}} }%
% \includegraphics[width=5.5cm]{SI_figures/SIFig1a.pdf} 
% \includegraphics[width=5.5cm]{SI_figures/SIFig1b.pdf}
% \includegraphics[width=9.0cm]{SI_figures/SIFig1c.pdf}
\includegraphics[width=0.9\textwidth]{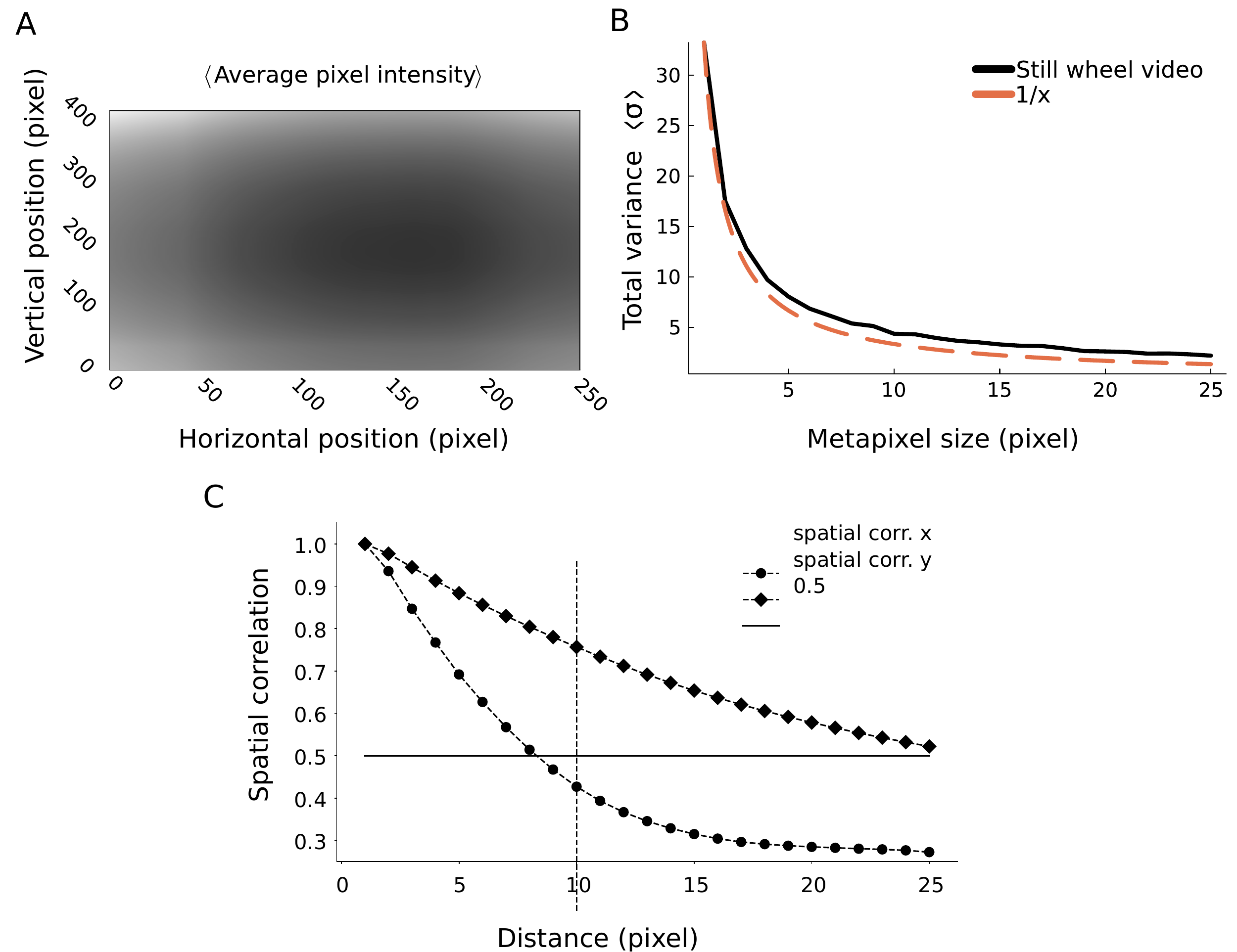} 

	\captionsetup{labelformat=empty}
	\caption{Figure S\ref{variance}:   A) Normalization matrix obtained from the external product of intensities averaged over rows and columns (see the text for the specific procedure).  B) Total time variance for $\omega=0$, as a function of the metapixel size $S$; C) Spatial correlation functions of the intensity pattern for $\omega=0$ on the $x$ and $y$ directions.}
%	\caption{\textit{a}) %Average beam 
%		Reference intensity matrix used to rescale pixel intensities.
%		%averaged over column and rows to minimize the distortion due to bright pixels.
%			\textit{b}) Trend of the variance of the metapixel intensity with still plate;  The average over the whole space-time matrix decays slowly when the metapixel size is above $S=10$ pixel. The variance is compared with a decaying characteristic length of \num{10} metapixel (dashed line).} 
	\label{variance}
\end{figure}

\paragraph{Scaling of pixel intensities.}
A video of duration  \SI{11}{\minute} has been recorded for different speeds of the rotating plate. Each video is encoded as a three-dimensional matrix whose indexes span the two spatial dimensions and the time. This way, the time evolution of the interference pattern is contained in a three-dimensional array of dimensions (250, 420, 9900) whose entries represent the back-scattered light intensity recorded at a pixel $(x,z)$ at a time $t$. 
Because the beam intensity is not uniform over the channel wall, one could normalize data with the average intensity of each pixel. However, the presence of very bright spots in the interference pattern makes this kind of normalization ineffective.
%  For this purpose, we averaged the intensity over rows and columns and computed each pixel intensity through the external product of these vectors. Leveraging the system's radial symmetry and obtaining an effective normalization matrix. The normalization matrix is shown in Fig.~\ref{variance}\textit{A}.
Thus we proceeded by taking a reference value at each pixel which has been computed in the following way. First, we averaged all the intensities over time obtaining a two-dimensional array (250, 420). Then, we performed averages of this array over rows and columns obtaining two one-dimensional arrays. Finally, the normalization matrix was obtained through the external product of such vectors and taking the square root of each entry (see Fig.~S\ref{variance}A).    

%averaging intensities over rows and columns and computing the external product of the two resulting vectors.  Leveraging the system's symmetry an effective normalization matrix is so obtained, as shown in .

\paragraph{Spatial coarse-graining and choice of the metapixel.}

The intensity data collected at every pixel have been averaged in space over squares of side $S=$ \num{10} pixels, the metapixel (MP). 
To determine the optimal  size, a set of \SI{5}{\minute} long data has been collected  with the plate at rest and the following steps followed:
\begin{itemize}
	\item[i)] Coarse-graining was performed on each frame by averaging intensities over the pixels of a square block of size $S$. At the beginning $S=$\num{1};
	\item[ii)] for each block, the variance $\sigma^2$ over all the frames of the video was computed;
 \item[iii)] The computed variances were averaged over the blocks of the entire interference image, yielding a value $\langle \sigma^2(S) \rangle$ for each choice of $S$
 \item[iv)] 	the block size $S$ was increased by one and the procedure repeated.
\end{itemize}
Notice that in the absence of noise on the detector, each variance should be strictly zero since we are considering the case with the plate at rest. Thus, the average variance $\langle \sigma^2(S) \rangle$ indicates the amount of background noise; being uncorrelated, this noise is expected to decrease for increasing sizes of the metapixel $S$. The results obtained in our setup are shown in Fig.~S\ref{variance}B). 
%The plot shows that after an initial quick drop, a slower decrease follows. Based on this trend, the size of the metapixel has been set where the decrease slows down, $S=10$, in the idea that a further increase of the metapixel size would result in a loss of information.  (Tab. S\ref{tab}).
Panel B shows that the standard deviation decreases as $S^{-1}$ as expected. However, the noise reduction obtained in this way is paid for by a loss of spatial resolution. Indeed, once we construct an interference image with metapixels of size $S$ we cannot resolve anymore any spatial modulation of the intensity over scales $<S$. This is a crucial point since we are interested in the characterization of the shear profile.  The criterion we used to stop the coarse-graining procedure is to choose an optimal $S$ which is smaller than the typical spatial correlation length of the intensity pattern in the video with the plate at rest. In this way, we minimize the resolution loss since each metapixel contains on average only correlated pixels which bring the same local information.  This procedure allows us to only filter the noise around local intensity values without losing information on spatial intensity modulations. In Fig.~S\ref{variance}C we show the spatial correlation functions of the intensity pattern for $\omega=0$ $s^{-1}$. We compared the ones obtained considering distances over both $x$ and $y$,  computed on single frames and then averaged over 100 frames. From this plot, it is seen that a reasonable choice for the metapixel size is $S=10$ since intensities within this distance are correlated in both directions.

%referring to the upper left panel of Fig. 3 of the main text(intensity pattern for $\omega=0$ $s^{-1}$), we want our metapixel to be as large as possible (in order to reduce the noise) but without overcoming the typical sizes of the light blobs (in order to maintain spatial resolution).

\section*{Autocorrelation function}

For each value of the plate rotational speed, we computed the connected time autocorrelation function for the metapixel centred in $(x,z)$ (horizontal and vertical coordinates):
\begin{equation}
g(x,z,t)=\frac{\langle I(x,z,t_0) I(x,z,t_0+t) \rangle_{t_0} - \langle I(x,z,t_0)\rangle^{2}_{t_0}}
{\langle I^2(x,z,t_0)\rangle_{t_0} - \langle I(x,z,t_0) \rangle^{2}_{t_0}}
\end{equation}
 %estimated over discrete times $t_m= m\Delta t$   as APL
 averaged over discrete initial times $t_m= m\Delta t$ as
\begin{eqnarray}
%\label
\langle I(x,z,t_0) I(x,z,t_0+t) \rangle _{t_0}
%_{t_m} APL
= \frac{\sum_{m=1,N_t} I(x,z,t_m+t) I(x,z,t_m)}{N}
\\ \nonumber
\langle I^2(x,z,t_0)\rangle_{t_0}= \sum_{m=1,N}I^2(x,z,t_m)/N_t\\
%N_t APL
\langle I(x,z,t_0) \rangle^{2}_{t_0}= \sum_{m=1,N}I(x,z,t_m)/N,
%N_t APL
\end{eqnarray}
\noindent with  $\Delta t= 1/15$, $N=9.9\cdot10^3$, corresponding to a time window of \SI{11}{\minute}, and \textbf{$N_t=N-\lceil t/\Delta t \rceil$}.
%\textbf{$N_t=N-t_m$} APL

The number of operations required for computing the autocorrelation function grows at least as $N \log N$ if it is done via a fast Fourier transform.  In the present case, it was preferred a straight computation, even if it grows as $N^2$, in order to avoid other drawbacks related to the calculation of the Fourier transform of a finite series.  To this aim, we developed a Julia package named \textit{interference.jl} described below. 
The autocorrelation function is then obtained by first averaging the intensities over each metapixel, thus reducing to a matrix of dimension (25,42,9900), and then as direct products of 3-d matrices (see Sec.~6).  
Being the system rotationally invariant, the $g(x,z,t)$ have been averaged over $x$ ($x \in [1,25]$),  resulting in the $g(z,t)$, reported in the main text.

\section*{Smoothing and fitting procedure}
%Considering 
Using 
raw intensity matrices over time, we often observed a sharp jump between the first and the second point of the autocorrelation function for some metapixels, 
suggesting the presence of some residual noise faster than our acquisition rate (15 fps).
%This behavior suggests the presence of some spurious signals with characteristic times faster than our acquisition rate (15 fps).  
In order to address this issue, the inverse times $\nu_C=\tau_C^{-1}$, shown in Fig. 5, have been estimated by fitting with the positive part of a Gaussian $\propto \exp{(-2t^2/\tau\tau_C)}$ the short-time regime of autocorrelation functions obtained from smoothed intensity signals. Here $\tau_C$ is the characteristic physical time related to the %intensity modulation
autocorrelation decay
 while $\tau$ is the time window duration of the running average used to smooth the signals. In the following, we provide a simple mathematical argument to justify this procedure.

The smoothed intensities can be expressed as
\begin{equation}\label{eq::smootheSig}
   I_s(t)=\frac{1}{\tau}\int_t^{t+\tau}dt'I(t')
\end{equation}
and the non-normalized smoothed autocorrelations read
\begin{equation}\label{eq::smootheAu}
\tilde g_s(t)=\frac{1}{T}\int_0^Tdt'I_s(t')I_s(t'+t).
\end{equation}
Here we assume $I(t)$ to be the signals with the time averages subtracted. To simplify it, we avoid including the spatial and the $\omega$ dependence in the notation.
%It is now necessary to 
Let us now 
show analytically how this smoothing suppresses the fast-decaying spurious contribution and why using a Gaussian fit for $\tilde g_s(t)$ is reasonable.
Using Eq. \eqref{eq::smootheSig} we can write:
\begin{equation}\label{eq::passages}
\begin{split}
\tilde g_s(t)=\frac{1}{T}\frac{1}{\tau^2}\int_0^T dt'\int_{t'}^{t'+\tau}dt''I(t'')\int_{t'+t}^{t'+t+\tau}dt'''I(t''') \\ = \frac{1}{\tau^2}\int_0^\tau dt_1\int_0^\tau dt_2 \frac{1}{T}\int_0^T dt'I(t_1+t')I(t_2+t'+t)
\end{split}
\end{equation}
where we changed variables as $t_1=t''-t'$ and $t_2=t''' - t' -t$.
The last integral of Eq. \eqref{eq::passages} can be expressed in terms of the original (i.e. non smoothed) autocorrelation function $\tilde g(t)$ only if the argument of one of the two integrated intensities is larger than the other. Indeed we have that 
\begin{equation}
\frac{1}{T}\int_0^T dt'I(t_1+t')I(t_2+t'+t)=\begin{cases} \tilde g(t_1-t_2-t) \quad \text{if} \quad t_1\ge t_2+t \\ \tilde g(t_2+t-t_1) \quad \text{if} \quad t_1 \le t_2+t \end{cases} \quad \text{with} \quad t>0
\end{equation}
From the above equation we 
%realize 
see that the smoothed autocorrelation can be written as a sum of two contributions:
\begin{equation}\label{eq::gsSplitted}
 \tilde g_s(t)=\frac{1}{\tau^2}\int_0^\tau dt_2\int_{t_2+t}^\tau dt_1 \tilde g(t_1-t_2-t)   + \frac{1}{\tau^2}\int_0^\tau dt_1\int_{t_1-t}^\tau dt_2 \tilde g(t_2+t-t_1) 
\end{equation} 
For $t \le \tau$, which is representative of the short-time decay of $\tilde g_s$, we have to consider both contributions while for $t \ge \tau$ only the second one holds. This is because $t_1 \in [0,\tau]$ so if $t \ge \tau$ the inequality $t_1\ge t_2+t$ is never true. 

In order to go on with the calculations, we need to assume a functional form for $\tilde g(t)$. The simplest choice is taking exponentially decaying correlations, with a characteristic time $\tau_C$: $\tilde g(t)=A\exp(-t/\tau_C)$. It coincides with modelling the intensity signals $I(t)$ as Ornstein-Uhlenbeck processes. We %now 
make this assumption and later verify \emph{a posteriori} his compatibility with empirical observations. By substituting exponential autocorrelations into Eq. \eqref{eq::gsSplitted} we get
\begin{equation}\label{eq::finalAutocorr}
\tilde g_s(t)=2A\frac{\tau_C}{\tau}\left[1 - \frac{\tau_C}{\tau}\left(1-e^{-\tau/\tau_C} \right)\cosh (t/\tau_C) \right] \quad \text{for} \quad t<\tau.
\end{equation}
The first thing to note is that the above expression is proportional to $\tau_C/\tau$ when this ratio is small. Thus, if the total signal contains a statistically independent fast spurious component with a characteristic time $\tau_C' \ll \tau \ll \tau_C$, then the autocorrelation of the smoothed signal will be dominated by the component proportional to $\tau_C/\tau$. 
Another important feature of Eq. \eqref{eq::finalAutocorr} is that it presents a zero first derivative and a negative second derivative for $t\sim 0$, consistent with the behavior observed in Fig. 4 of the main text (compare panel C1 and C2).
Considering now the normalized autocorrelation function $g_s(t)=\tilde g_s(t)/\tilde g_s(0)$ and expanding it at the leading order we find:
\begin{equation}
\tilde g_s(t)=1-\frac{2t^2}{\tau\tau_C}. \label{eq::gsExpansion}
\end{equation}
which is equivalent to the second-order expansion of the Gaussian $\exp(-(t/\tau_G)^2)$, 
where $\tau_G=\sqrt{\tau\tau_C/2}$. Such result %explains why we 
justifies the 
use a Gaussian to fit  the short-time part of the experimental autocorrelation function and %justify 
explains
how $\tau$ and $\tau_C$ appear in it. In principle %we should have performed 
one could perform the fit directly with 
Eq. \eqref{eq::finalAutocorr}, but we have found more efficient the use of a 
Gaussian since it properly %decays towards nearly interpolating 
interpolates the $t\sim\tau$ 
regime of Eq. \eqref{eq::gsSplitted} where Eq. \eqref{eq::finalAutocorr} ceases to be valid.

\section*{Characteristic time and shear band limit detection}

For each considered plate velocity, $\omega_p$, the characteristic decorrelation time at each depth $z$, $\tau_C(z,\omega_p)$,  has been determined with two methods, obtaining compatible values. 

\subsection*{Gaussian fit}

The method that yields the most accurate results is analyzing the autocorrelation function of the smoothed intensity signal. The short-time decay in the autocorrelation is fit with the positive part of a Gaussian function (coherent with Eq. \ref{eq::gsExpansion}), which allows extracting the characteristic time as a function of the metapixel height, $\tau_C(z)$, for each of the eight plate speeds considered. %
In the analysis, the
%The
 inverse of the characteristic time $\nu_C(z)$ has been
 % was also computed, and it is
  used %in the analysis 
  because % it presents 
  less noisy.
  %than $\tau_C(z)$ itself.

We have used a least-square method from the module {\em LsqFit} of Julia programming language to fit the Gaussian. The code implements the Levenber-Marquardt algorithm for non-linear fitting. The fitting function is a Gaussian with only the characteristic time $\tau_G=\sqrt{\tau\tau_C/2}$ as a parameter:
\begin{equation}
	G(x,\tau_C, \tau) = \exp \left[ -2 \left( \frac{x}{\sqrt{\tau_C\tau} }\right)^2 \right]\label{eq::gaussian_fit}
\end{equation}
\noindent where the $\tau $ is 
the smoothing time, as discussed in the previous section,  set equal to 5 (equivalent to $1/3$ s).
%which
This allows extracting the characteristic $\tau_C(z)$, for each of the eight plate speeds considered. Its inverse, $\nu_C(z)$, expresses the  decorrelation rate  and has been employed in the analysis

\begin{figure}[h]
	\centering
	\includegraphics[width=0.65\linewidth]{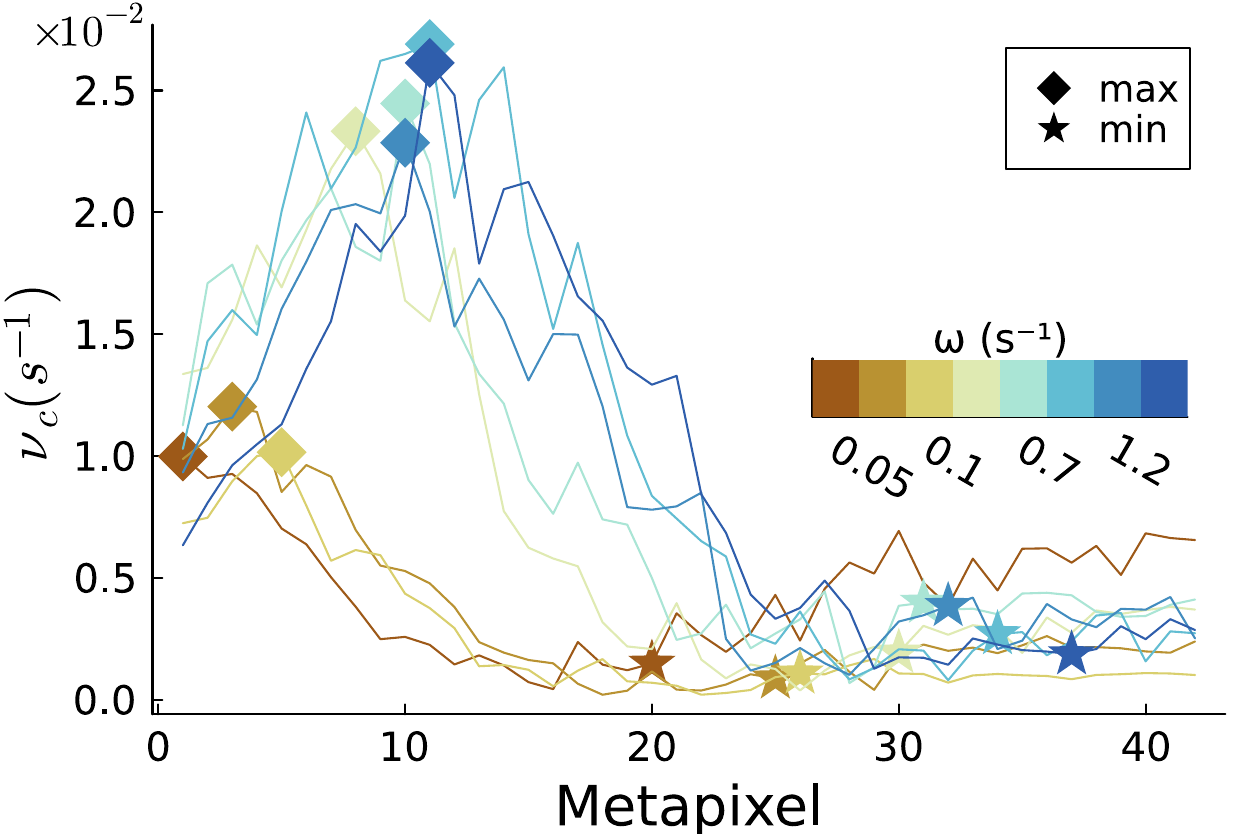}

			\captionsetup{labelformat=empty}
	\caption{Figure S\ref{fig:shearbanddetection}: %\textcolor{red}{A) UNA PROVA.} 
 Profiles of the characteristic time for all the plate velocities considered. The minima and maxima of the inverse characteristic time are shown as stars and diamonds, respectively. The shear band is defined as the region between the minimum and the first maximum. The data shown here are the same used for Fig. 6 of the main text.} 
	\label{fig:shearbanddetection}
\end{figure}

\paragraph{Shear band detection.} The % inverse characteristic time
  decorrelation rates $\nu_C(z)$ allow to individuate the shear band %and 
characterizing the  quasi solid and liquid phases. 
%The shear band is defined as the region between the minimum $\tau_C^{min}$ (the maximum of $\nu_C(z)$) and the first maximum $\tau_C^{max}$ (the minimum of $\nu_C(z)$). 
As seen in Fig. 5 of the main text, 
following the first maximum, %the inverse characteristic time 
$\nu_C(z)$ first decays and then 
enters a plateau associated with the solid phase. We assume the shear band as the region between the  maximum of $\nu_C(z)$ and the beginning of the plateau.
The starting of the shear band is easy to localize.
%we simply considered the depth with the maximum inverse characteristic time. 
Conversely, to individuate the plateau 
%finding the minimum was 
is less straightforward because of the noise in the data. We 
% exploited the same noise  to
determine, within a  certain approximation, the beginning of the plateau by associating it with the first relative minimum of  $\nu_C(z)$ after the maximum. %We used a simple algorithm to find the minimum described in the following. 
Assuming the curve to be smooth, to find the first minimum,  %we should 
one could compute the finite differences of the decorrelation rates $\Delta \nu_C(z) = \nu_C(z+1) - \nu_C(z)$, and look for the first point where the difference is non-negative. However, the data is noisy, and the finite differences can be non-negative also before the actual minimum in the data. Hence, we averaged the finite differences over a window of size $w$, and looked for the first point where the average is positive. A %fair 
good estimate is obtained by averaging over all the window lengths in the range \SIrange{3}{13} {\text{metapixels}}. The maxima and the minima individuated with this method are shown in Fig. S\ref{fig:shearbanddetection}, together with the corresponding $\nu_C(z)$ profiles. The error bars in Fig 8 of the main text represent the standard deviation across the seven window lengths considered.

\subsection*{Half-height method}
\begin{figure}[htbp]
	\centering
	\includegraphics[width=1.\linewidth]{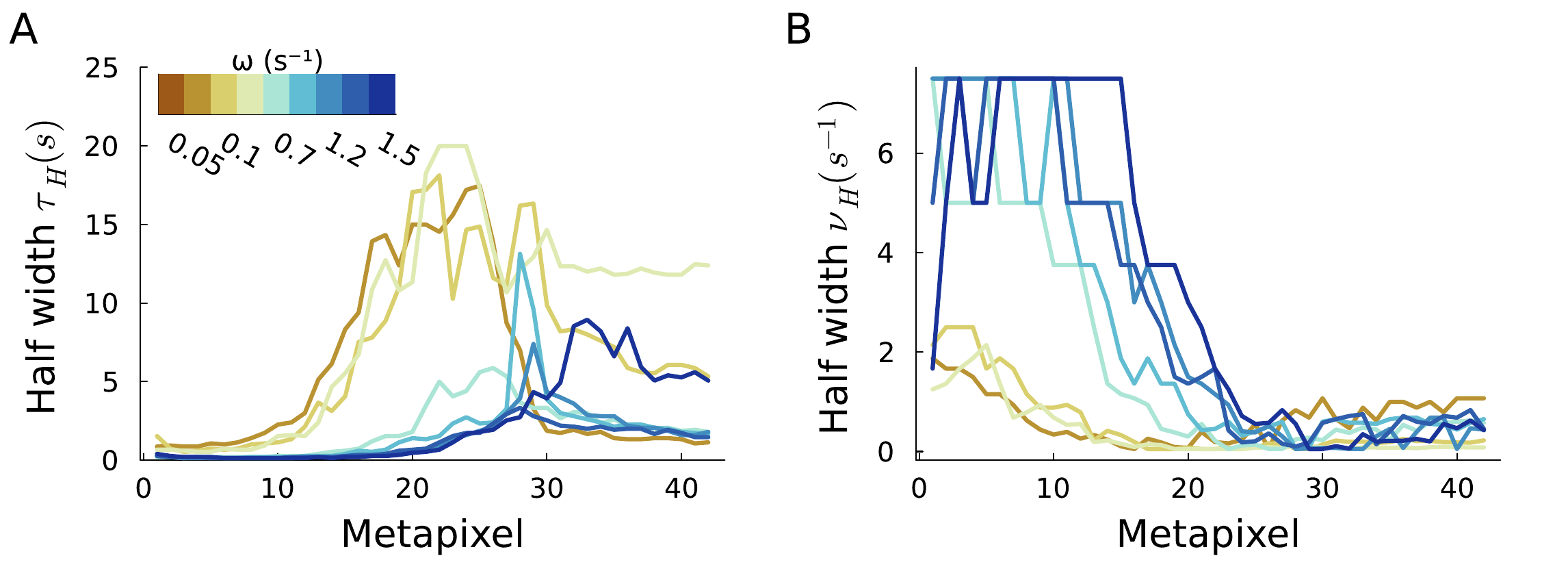}
		\captionsetup{labelformat=empty}
	\caption{Figure S\ref{fig:tauH}: Characteristic time (and its reciprocal in the right panel) computed with the half-height method. The right panel better shows the saturation of the method in the short timescales corresponding to the start of the shear-band.} 
	\label{fig:tauH}
\end{figure}

A second method to analyze the data was also tested; this method requires less analytical insight into the smoothing procedure and can be performed without fitting the auto-correlation decays.
To estimate the characteristic time, we used the half-height method. The characteristic time $\tau_H(z)$ is when the (raw) auto-correlation function decays to half its initial value. The half-height method is less accurate than the Gaussian fit and more sensitive to noise; however, it can be performed on non-smoothed data and is therefore simpler and virtually unbiased by any timescale introduced for the analysis.
The profiles of $\tau_H(z)$ and their inverse $\nu_H(z)$ are shown in Fig. S\ref{fig:tauH}. Because of the relatively coarse temporal resolution (determined by the camera frame rate), the resolution of the half-height method saturates on the short timescales associated with the beginning of the shear band. Consequently, the profile of $\nu_H(z)$ has a plateau near the maximum, and it is impossible to precisely detect the upward shear band limits. The right panel in Fig. S\ref{fig:tauH} illustrates this limitation.

\begin{figure}[h]
	\centering
	\includegraphics[width=1.\linewidth]{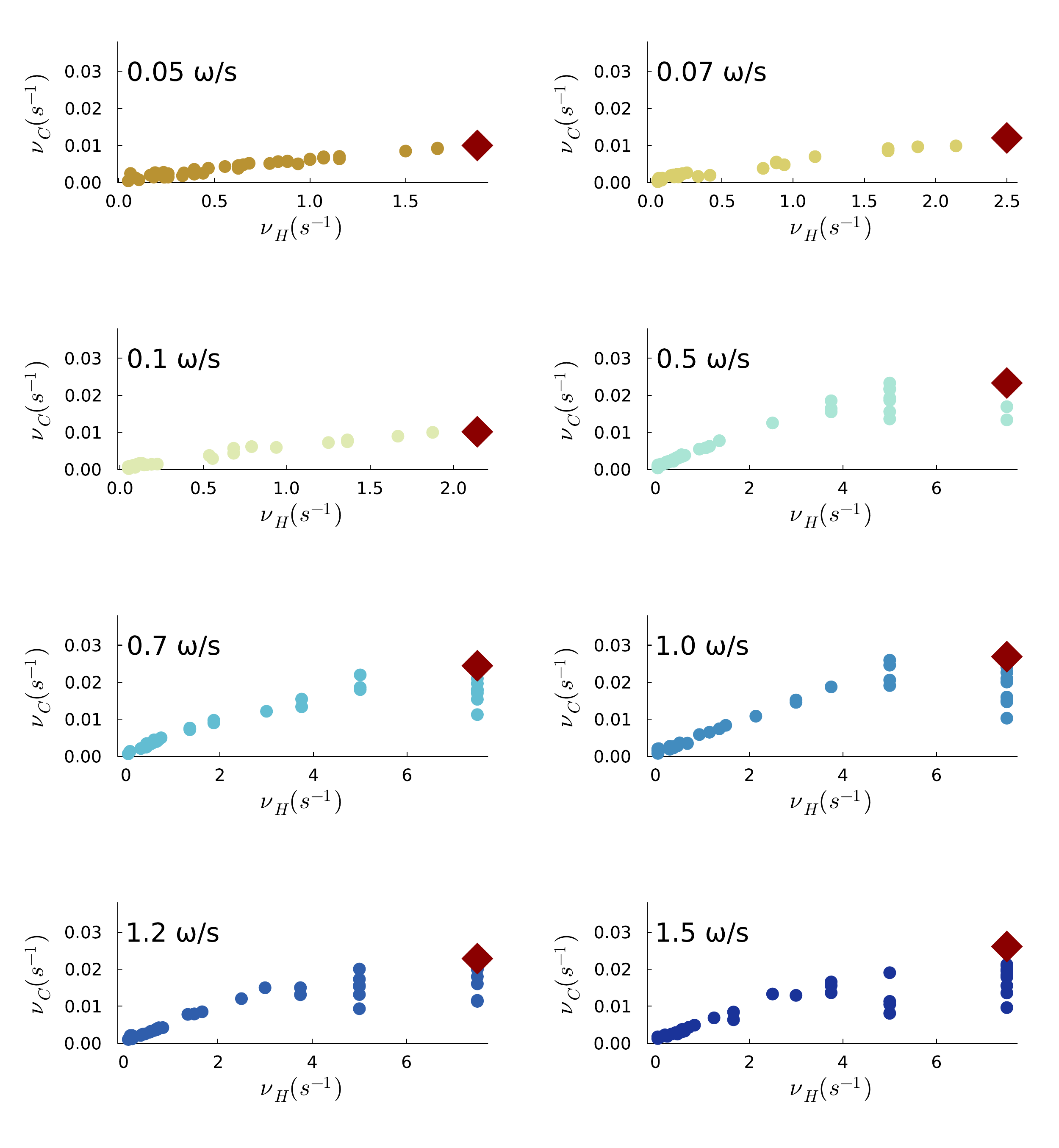}
			\captionsetup{labelformat=empty}
	\caption{Figure S\ref{fig:tauComparison}: Comparison of the decorrelation rate computed with the Gaussian fit $\nu_C$, and the one computed with the half-height method $\nu_H$. The two methods yield compatible results. The diamond indicates the maximum value of the decorrelation rates, which is used to define the starting of the shear band. For large shear rates,   $\nu_H$ saturates (see Fig. S\ref{fig:tauH}B) }
	\label{fig:tauComparison}
\end{figure}

However, it is possible to compare the two methods and verify whether they yield compatible results. The scatter plots in Fig. \ref{fig:tauComparison} illustrate the characteristic  times and their inverse for each metapixel and each plate velocity, showing that there is good agreement between the two measures.

%{\color{red}NB: le label orizzontali credo siano $\nu_C$ e non $\nu_H$; le label verticali delle fig in alto appaiono  incomplete; le stelle nella Fig. S£ indicano i minimi e on i massimi, sarebbe meglio omogeneizzare}
\section*{Code for interference image  analysis}

The code \textit{interference.jl} has been developed for the experiment itself and is available on GitHub\footnote{\url{https://github.com/aquaresima/interference.jl}}. 
The software uses bindings to Python Numpy to transform the \verb+.avi+ recordings into \verb+.npy+ matrices, which are then imported in Julia. We preferred to use Julia because the computation of the coarse-graining and the correlation is faster. The original matrices are sized \numproduct{250x420} 
$\times$ T, where T is not fixed. 
%, is different for each speed. 
The first two dimensions are the corresponding  width and height pixel number of the single frame, the third dimension is the number of frames. Each element of the matrix represents the intensity at the corresponding pixel encoded in 8 bits - thus  spanning from 0 to 255. 

%\begin{figure}[h]
%	\centering
%	\includegraphics[width=1.\linewidth]{SIFig6}
%		\captionsetup{labelformat=empty}
%	\caption{Figure S\ref{fig:tauComparison}: Algorithm for detection of the shear-band, see text} 
%	\label{fig:tauComparison}
%\end{figure}

The code \textit{interference.jl}  offers a set of tools for data processing.

\paragraph{Signal coarse-graining and autocorrelation computation.}

The matrices are coarsened in time and space with a non-overlapping average-pooling algorithm - i.e. the value of coarsen matrix is the average of pixels' intensity in the corresponding region of the fine grain matrix.
The spatial coarsen reduces raw signal noise, the coarsen in time also removes uninformative correlations on time scales that are irrelevant for the present study. The choice of the coarsening size is left to the user. 
The coarsen matrices are then stored in \verb+Float32+ format to limit the memory occupation.

From the obtained three-dimensional matrix, the autocorrelation functions are then computed as direct products of 3D matrices realized in a parallelized way. 

\paragraph{Estimation of the correlation time}

The time correlation function is computed at each metapixel within the assumption of stationary dynamics. 
Being $v_{ij}(t) = v_{ijt}$ the intensity at each metapixel,
the correlation algorithm returns a tridimensional matrix:
\begin{equation}
\hat C_{i,j,s} =
 T^{-1}\sum_{t=0}^{T} 
 \frac{v_{ij}(t)v_{ij}(t+s)-T^{-1}\sum_{t=0}^{T}v_{ij}^2(t)}
{T^{-1}\sum_{t=0}^{T}v_{ij}^2(t) -(T^{-1}\sum_{t=0}^{T}v_{ij}(t))^2}.
\end{equation}
The time dimension \textit{T} -the maximum correlation time- is determined by the available number of frames $N_f$ and the maximum lag $s$: $T=N_f-s$. 
%been chosen such $C_{i,j,T} \simeq 0$ $\forall i,j$. 
%Thus, the matrix is equivalent to the scalar-functions set, one for each metapixel's position \textit{i,j} dependent on the discrete-time \textit{s}. 
Thus, each matrix element corresponds to the intensity autocorrelation at the metapixel $i,j$ at the discrete time $s$.
Assuming %rotiational
translational symmetry on the $x$-axis, the correlation function is averaged over the horizontal coordinate $i$ to obtain a 2-d matrix of scalar functions of $s$, $\hat C_{j,s}$.

The characteristic decorrelation time $\tau_j$  is then determined  with two methods: 
\begin{itemize}
\item \textit{Half-height characteristic time}: A decaying characteristic time scale of each height is determined as the value of $s$  at which the function assumes half of its initial value (i.e. 0.5). One is then left with a vector %2-d matrix 
of characteristic times: $\hat \tau_{j}$.

\item \textit{Gaussian characteristic time}: The first 10 points of the correlation function are fitted with a Gaussian function. The characteristic time is then computed as described in Eqs. (\ref{eq::gaussian_fit}).
\end{itemize}

\paragraph{Shear band detection.}
\hfill\break

The detection of the jump between solid and fluid state, respectively left and right in Fig.  S\ref{fig:shearbanddetection}, corresponds to detect the maximum and first minimum of the function $\tau_i$. The algorithm described in Sec. {\it Characteristic time and shear band limit detection} is implemented in the code.

\end{document}